\documentclass[
 amsmath,amssymb,
 twocolumn,
 aps,
 pre,
]{revtex4-2}

\usepackage{graphicx}
\usepackage{dcolumn}
\usepackage{bm}
\usepackage{xcolor}
\usepackage{amsmath}
\usepackage{multirow}
\usepackage{soul}

\begin{document}


\title{Yielding in amorphous solids reveals an age-dependent intrinsic lengthscale}
\author{Aparna Sreekumari}
\affiliation{Department of Physics, Indian Institute of Technology Palakkad, Nila Campus, Kanjikkode, Palakkad 678623, Kerala, India.}
\author{Monoj Adhikari}
\affiliation{Heinrich Heine University Düsseldorf, Universitätsstraße 1, 40225 Düsseldorf, Germany.}
\author{Nandlal Pingua}
\affiliation{National Institute of Technology Tiruchirappalli 620015, Tamil Nadu, India.}
\author{Vishnu V. Krishnan}
\affiliation{University of Tokyo, 7 Chome-3-1 Hongo, Bunkyo City, Tokyo 113-8654, Japan.}
\author{Shilditya Sengupta}
\affiliation{Department of Physics, Indian Institute of Technology Roorkee 247667, Uttarakhand, India.}
\author{Pinaki Chaudhuri} 
\affiliation{The Institute of Mathematical Sciences, CIT Campus, Taramani, Chennai 600113, India.}
\altaffiliation[Also at ]{Homi Bhabha National Institute, Anushaktinagar, Mumbai 400094, India.}
\author{Smarajit Karmakar}
\email{smarajit@tifrh.res.in}
\affiliation{Tata Institute of Fundamental Research, 36/P, Gopanpally Village, Serilingampally Mandal,Ranga Reddy District, Hyderabad, 500046, Telangana, India.}
\author{Vishwas V. Vasisht}
\email{vishwas@iitpkd.ac.in}
\affiliation{Department of Physics, Indian Institute of Technology Palakkad, Nila Campus, Kanjikkode, Palakkad
 678623, Kerala, India}

\date{\today}

\begin{abstract}
Understanding how amorphous solids yield under shear is central to predicting material failure, yet prescribing reliable local yielding criteria remains a fundamental challenge. Here, through a mesoscale analysis of localized yielding, we reveal an intrinsic length scale ($\zeta$) that governs local failure, and demonstrate that $\zeta$ grows with the age of the system. The age dependence shows up not only in the features of the distribution of local yield stress but also in the pseudogap exponent $\theta$, which provides a measure of marginal stability of the amorphous solids. These insights are made possible by a new method—termed the soft matrix approach—that allows local regions of an amorphous solid to yield within a minimally constrained, elastically coupled environment. By overcoming key limitations of earlier techniques, our approach provides a robust platform for probing failure mechanisms, particularly in soft disordered materials and paves the way for improved elastoplastic modeling of disordered solids.
\end{abstract}

\maketitle


\section{\label{secIntro}Introduction}
{A}morphous solids are disordered materials lacking long-range structural order and include a wide range of everyday substances, such as colloids, foams, emulsions, granular media, and metallic and silicate glasses \cite{bonn2017yield}. Under external load such as shear deformation, they initially respond elastically and yield beyond a threshold strain before reaching steady-state flow \cite{varnik2004study, fielding2014shear, bonn2017yield, 2018Nicolasg}. Unlike crystals, the microscopic shear response of amorphous materials still lacks a complete understanding. The absence of crystalline order makes it difficult to identify the carriers of plasticity. Over the past two decades, theoretical and experimental studies have shown that plasticity in these spatially heterogeneous systems proceeds via localized non-affine rearrangements that redistribute elastic stress \cite{argon1979plastic, falk1998dynamics, maloney2006amorphous, berthier2025yielding}. Complementing these findings, simulation studies of athermal systems under quasistatic shear have revealed displacement fields with quadrupolar symmetry, consistent with Eshelby-type inclusions \cite{eshelby1957determination, maloney2006amorphous, tanguy2006plastic, jensen2014local}.

Amorphous materials exhibit pronounced spatial heterogeneities in both structural and dynamical properties \cite{2015KarmakarRPP}, which are central to understanding their complex response under deformation. In particular, recent efforts have focused on identifying a characteristic length scale that governs relaxation in driven systems, due to its relevance for shear start-up \cite{fielding2014shear, jaiswal_PRL_2016, shrivastav2016yielding, divoux2016shear, popovic2018elastoplastic, vasisht2020emergence}, yielding \cite{Ozawa_PNAS_2018, barlow2020ductile, leishangthem2017yielding, sastry2021models}, and flow heterogeneities \cite{dhont1999constitutive, schall2010shear, fielding2014shear, divoux2016shear, vasisht2018permanent}. Static correlation lengths have been extracted using non-affine deformation protocols in colloidal glasses and supercooled liquids \cite{weeks2007short, 2012MosayebiJCP, 2014MosayebiPRL}, though these typically reflect shear-induced structural correlations. Similar methods applied to athermal colloidal suspensions under steady-state flow reveal dependence on both strain and shear rate \cite{vasisht2018rate}. However, direct comparison with shear-free static length measures—such as the point-to-set length \cite{2008BiroliNPhys}, finite-size scaling of relaxation times \cite{2009KarmakarPNAS}, or minimum eigenvalues \cite{2012KarmakarPhysicaA} remains lacking.

Despite the absence of a clear definition for the characteristic length scale, several approaches have been proposed to estimate the size of plastic events or the number of particles involved in non-affine rearrangements. These include analyzing structural features (e.g., free volume \cite{spaepen1977microscopic}, local ordering \cite{jack2014information, taiki2017avalanche}) and linear response metrics (e.g., elastic moduli \cite{2009TsamadosPRE}, soft modes \cite{rottler2014predicting}). Among these, the {\it frozen matrix} method has gained widespread use for computing local mechanical properties. Originally proposed by Sollich \cite{Sollich2010}, this method allows only a designated  region to relax independently under an externally applied affine deformation, while the rest of the system is frozen—suppressing plastic events outside the chosen region. The stress response within the target region to the imposed strain is used to infer its local mechanical properties. 

{\color{black} The frozen matrix method has been extensively applied to studies of amorphous solids, particularly to analyze yielding thresholds and viscoelastic responses \cite{Mizuno2013, Puosi2015, patinet2016connecting, barbot2018local, ruan2022predicting}. Typically, the size of the target region is empirically chosen to be large enough to capture the material’s linear response \cite{Mizuno2013, barbot2018local, ruan2022predicting}. However, the abrupt truncation of non-affine displacements at the boundary of the target region often leads to significant overestimation of the local yield stress and elastic modulus \cite{Mizuno2013, barbot2018local, 2020Vasisht}. To address these limitations, a refinement known as the {\it thawed matrix} method has recently been proposed \cite{rottler2023thawed}. This approach incorporates a buffer layer between the target region and the frozen background, reducing artifacts caused by sharp displacement truncation. With the buffer size as a tunable parameter, compared to the frozen matrix approach, the thawed matrix method provides a more reliable characterization of plastic events. However, it still falls short in capturing features related to the pseudogap, which captures the information related to local regions close to yielding and hence is central to understanding marginal stability of amorphous solids \cite{karmakar2010statistical, hentschel2011athermal, lin2014density, lin2016mean, lerner2018protocol}.}

In this study, we address key challenges in understanding yielding and plasticity in amorphous solids: (i) identifying the minimal size of a local region capable of sustaining an independent plastic failure, (ii) accurately computing local mechanical properties while suppressing background plastic events without introducing boundary artifacts, and (iii) capturing the age dependence of the local yield stress statistics. To this end, we introduce a novel approach, referred to as the "soft matrix" method, for probing local mechanical properties. Using this method, in different models of amorphous solids, we find that the local yield strain exhibits an exponential decay at smaller sub-system sizes - revealing an emergent length scale associated with local yielding. Beyond this length scale, the yield strain recovers the expected power-law behavior. The method also captures a clear dependence of this length scale on the sample’s age. By performing computationally intensive measurements of local yield stress thresholds, we show that the resulting statistics are well described by a Weibull distribution, with both the scale and shape parameters exhibiting distinct age dependence. At the lower end of the distribution, we observe a power-law tail, allowing us to extract the associated pseudogap exponent—which, notably, also varies with sample age. We demonstrate age-dependent pseudogap behavior, evident both in the small-threshold power-law regime of the yield threshold distribution and in the large length scale variation of the Weibull shape parameter.

\begin{figure}[t]
\centering
\includegraphics[width=0.95\linewidth]{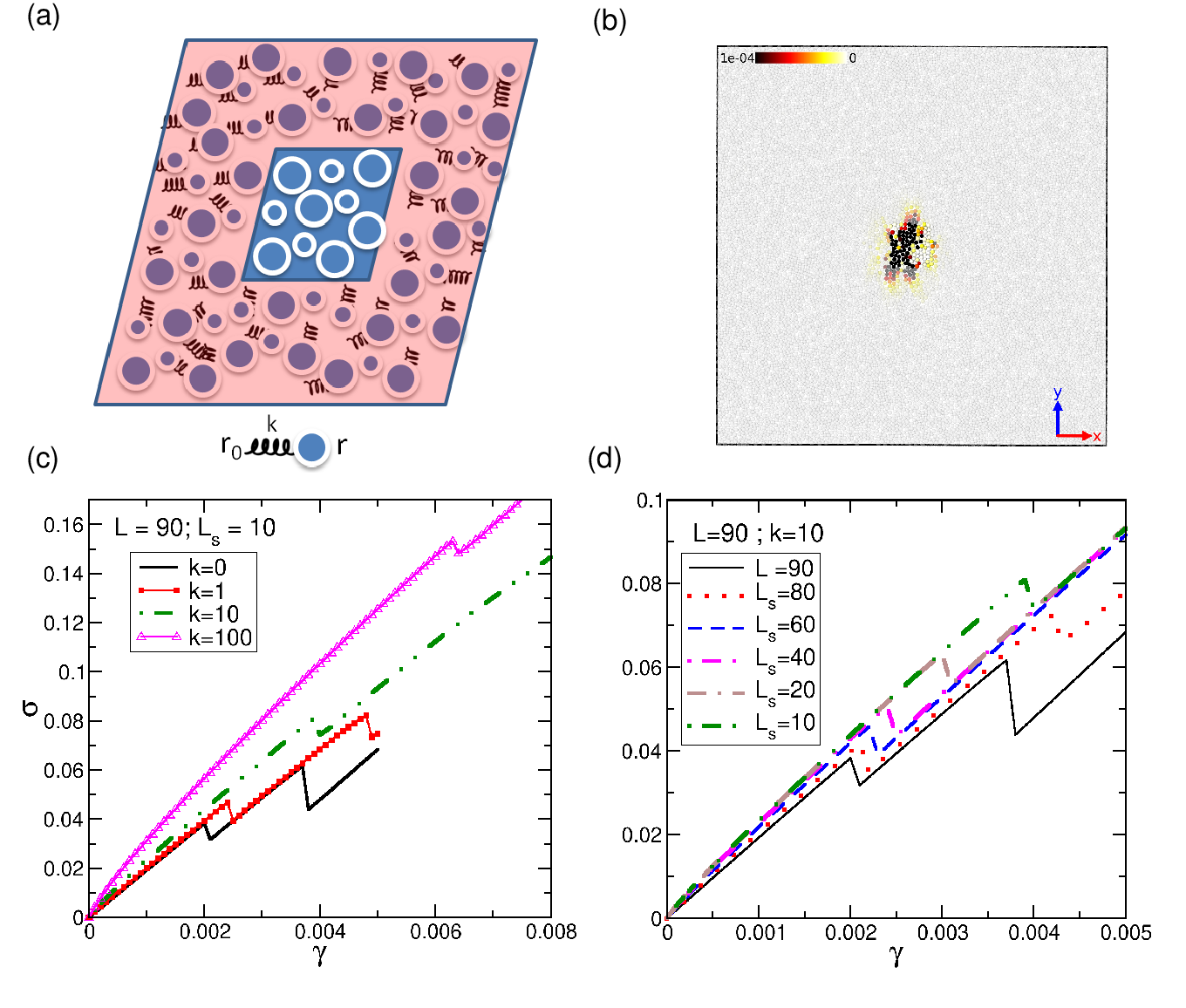}
\caption{{\bf Soft Matrix Methodology.} (a) Schematic representing the soft matrix method. The sub-system (centre blue box) is subjected to unconstrained relaxation, whereas the background (surrounding pink region) relaxes with a constraint imposed by the spring tether (see text for details). (b.) Snapshot from 2D BMLJ simulation shows a plastic event confined to the subsystem, even as the background is allowed to relax. Here, {\it x} is the shearing or flow direction, and {\it y} is the gradient direction. The coloring scheme is based on particle displacement, where white indicates no displacement and black corresponds to the largest (of magnitude 0.0001). (bottom) Variation of stress ($\sigma_{xy} \equiv \sigma$) as a function of strain ($\gamma$) for (c) fixed subsystem size ($L_s=10 a$) and varying $k$ value and (d) fixed $k$ value  ($k=10 \epsilon/a^2$) and varying subsystem size.}
\label{fig1}
\end{figure}

\begin{figure*}[htpb]
\centering
\includegraphics[width=0.95\linewidth]{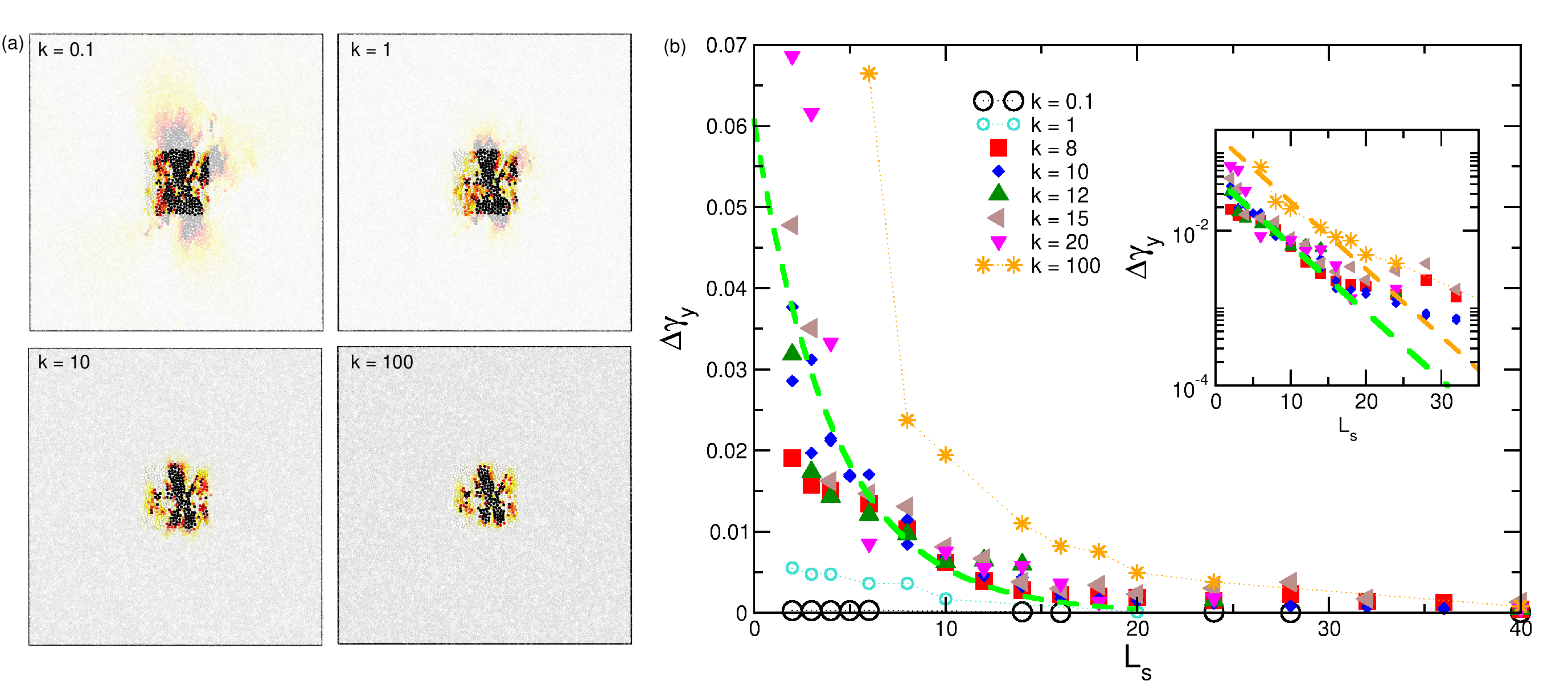}
\caption{{{\bf The k dependence.} (a) Displacement maps obtained from 2D BMLJ simulations for varying k values (0.1, 1, 10 and 100). The coloring scheme - heatmap is based on the particle displacements (white indicates no displacement, and black corresponds to the displacement of magnitude 0.0001). The region where the soft matrix is imposed is made translucent to emphasize the sub-system (with size $L_s = 20 a$). The top two panels (k = 0.1 and 1 $\epsilon/a^2$) are displayed with higher translucency than the lower ones (k = 10 and 100 $\epsilon/a^2$). The reduced translucency in the lower panels is intentional, as it highlights the displacements near the sub-system boundaries that are otherwise less visible. With the increasing k value, the displacement outside the probe region is suppressed in a gradual manner; notably, near the boundaries, the displacements are softened rather than sharply cut off. (b) Relative yield strain $\Delta \gamma_y$ as a function of sub-system size $L_s$ for varying k values, computed for a poorly aged sample. Inset of (b) shows the same plot in lin-log scale to highlight the exponential behavior. The dashed line corresponds to the exponential fit $\propto exp(-L_s/\zeta)$. }}
\label{fig6}
\end{figure*}

\section{\label{secSMmethod}Soft matrix method}
We introduce a soft matrix method, an improvement over the frozen matrix approach   \cite{Sollich2010, Mizuno2013, Puosi2015, patinet2016connecting, barbot2018local}, to study local yielding in amorphous solids. A local sub-region of characteristic size $L_s$ (i.e., an area $L_s^2$ in 2D or a volume $L_s^3$ in 3D) is chosen within a larger system of linear size $L$, with the rest treated as the background. Following the athermal quasi-static shear (AQS) protocol (see Methods), a small shear strain $\delta \gamma \approx 10^{-5}$ is applied, inducing an affine displacement across the entire system. The key feature of our method lies in the relaxation step: the sub-region is allowed to relaxed completely without constraints, while the background is restricted to affine relaxations only, preventing any plastic events. This is achieved by applying a harmonic restoring force $F_r = -k r$ to background particles, where $r$ is the displacement from their affinely deformed positions and $k$ is the spring constant, the only tunable parameter. This setup allows for controlled probing of local yielding while maintaining mechanical continuity with a background that permits affine relaxation without plastic events. We apply this method to measure the yield strain associated with the first plastic event in four amorphous solid models: 2D BMLJ, 3D soft-Rep, 3D $SiO_2$ and 3D metallic glass. Simulations were performed using the LAMMPS molecular dynamics package \cite{LAMMPS}. {\color{black} All quantities presenting in this work are expressed in simulation units (unless otherwise mentioned), where the unit of length is $a$, the average diameter of the particle, energy is $\varepsilon$, the energy scale in the interaction potential and the shear stress as well as the storage modulus are $\varepsilon/a^d$, where d is 2 or 3 depending on 2D or 3D system.}

In Fig. \ref{fig1} (a) we illustrate the method schematically. The blue sub-region is fully relaxed without constraints, while the surrounding pink background is subjected to an additional harmonic restoring force with a suitable spring constant 
$k$, ensuring that irreversible plastic motion is effectively suppressed in this region. In Fig. \ref{fig1} (b) we show a snapshot, at the first plastic event, from the  2D BMLJ simulation ($L=90a$, $N=10K$). The coloring scheme is based on particle displacement, where white indicates no displacement and black corresponds to the largest. The snapshot clearly shows that displacement is concentrated within the sub-region, with minimal to no motion in the background.

\begin{figure}[t]
\centering
\includegraphics[width=0.98\linewidth]{Fig2.eps}
\caption{{\bf Characteristic Length Scale.} Relative yield strain $\Delta \gamma_y$ as a function of sub-system size $L_s$ for four different systems, which are (a) 2D BMLJ (b) 3D soft-Rep (c) 3D SiO2 and (d) 3D metallic glass. The open symbols are data obtained from the soft matrix method. The {filled} symbols are from the frozen matrix method. Different symbols correspond to different system sizes. The thick dashed line is the exponential fit, and the thin dashed line is just a guide to the eye.}
\label{fig2}
\end{figure}

\section{\label{secSSR}The first plastic event - shear strain analysis}
For a given sample prepared using the standard quench protocol, we perform an AQS simulation of the unconstrained system to obtain the bulk stress-strain curve. From this, we identify the first local yielding event—referred to as the first plastic event—and determine its spatial location along with the participating particles (see {\it Methods}). Based on the identified yielding region, we define a sub-region of linear size $L_s$ centered on the event and restart the AQS simulation. In this setup, the particles within $L_s$ is allowed to relaxed without constraints, while the surrounding background is only allowed to undergo affine relaxation, suppressing non-affine displacements
In Fig. \ref{fig1} (c), we show the load curves (shear stress $\sigma_{xy} \equiv \sigma$ vs. applied strain $\gamma$) for a range of {\it k} values at a fixed sub-system size ($L_s=10 a$) for the 2D BMLJ model. The bulk load curve corresponds to the case $k=0 \epsilon/a^2$. Increasing the value of {\it k} enhances the rigidity of the background (with ${\it k \rightarrow \infty}$ recovering the frozen matrix limit) as reflected in a steeper slope of the stress-strain curve. Additionally, higher {\it k} values delay the first plastic event, shifting it to larger strain and increasing the corresponding local yield stress. {\color{black} In this work, we fix the {\it k} with an appropriate value to suppress background plastic events during stress relaxation, without the boundary effects. 
To determine the value of {\it k}, we monitor changes in bulk mechanical properties (e.g., pressure, elastic modulus) across a range of {\it k} and simultaneously track displacements in both the sub-region and the background. For a suitable {\it k}, the background exhibits only small, elastic displacements and the bulk mechanical properties are not significantly affected by the additional harmonic potential (see {\it SI}).} {In Fig. \ref{fig6} (a), we present the displacement maps (for 2D BMLJ) illustrating the microscopic effect of the soft matrix. Increasing k value gradually suppresses plastic events outside the sub-system, with correlations near the boundaries being softened rather than sharply cut off.}

Having fixed the $k$ value, we now examine the effect of sub-system size $L_s$ on the  load curve, specifically on the yielding strain. In Fig. \ref{fig1}(d) we show load curves for $\it k = 10 \epsilon/a^2$ for varying $L_s$. The solid black line represents the load curve for an unconstrained system with a local yield strain of $\gamma_y \approx 0.002$ at the first plastic event. With varying $L_s$, the initial linear response does not show much variation, but we observe a systematic change in local yield strain $\gamma_y$ and local yield stress $\sigma_y$. With the decrease in $L_s$, $\gamma_y$ and $\sigma_y$ increases. Approaching $L_s \rightarrow L$, we recover the unconstrained load curve ({{\it SI} provides $L_s$ dependent displacement map for 2D BMLJ.}). We find similar features in the other three model systems that we have studied. The value of $k$ was fixed for each of the systems according to the above-prescribed protocol. {\color{black} One can rationalise the increasing yield stress and strain with decreasing $L_s$ as a result of the background curtailing the motion of particles in the sub-system and hence increasing the barrier height for the first plastic event.} However, we find a more interesting observation when we look at the variation of $\gamma_y$ with $L_s$. 

\section{\label{secZeta}A characteristic length scale $\zeta$}
Next we study the variation of relative yield strain $\Delta \gamma_y$ with the sub-system size $L_s$ for all four different models of amorphous solids. The relative yield strain $\Delta \gamma_y$ is defined as the deviation of $\gamma_y$ from the bulk value: $\Delta \gamma_y = \gamma_y (L_s) - \gamma_y (L)$ where $\gamma_y (L)$ and $\gamma_y (L_s)$ are the local yield strain for the bulk ($k=0\epsilon/a^2$) and the sub-system respectively. {In Fig.~\ref{fig6}(b), we present $\Delta \gamma_y$ vs.\ $L_s$ for the 2D BMLJ (poorly aged sample; L=320a; N=125K) across a wide range of $k$ values. For a nearly unconstrained system ($k=0.1\epsilon/a^2$), the yield strain matches the bulk value, giving $\Delta \gamma_y \approx 0$. In the range $8 \epsilon/a^2 < k < 20 \epsilon/a^2$ (where bulk properties remain essentially unchanged - see {\it SI}), $\Delta \gamma_y$ decreases with $L_s$, and the yield strain values remain comparable. At $k=100 \epsilon/a^2$, the overall trend is similar, but noticeable deviations appear with bulk behavior showing significant changes. The inset of Fig.~\ref{fig6}(b) shows the same data on a lin-log scale, clearly revealing the exponential regime independent of k value. Having established the robustness of the exponential decay, we now turn to examining this behavior across four different systems: 2D BMLJ, 3D soft repulsive model, 3D SiO$_2$, and 3D metallic glass. In Fig.~\ref{fig2}, we show the $L_s$ dependence of the average $\Delta \gamma_y$, computed over multiple initial configurations and, where possible, different system sizes.} All four systems show  $\Delta \gamma_y$ decrease exponentially with an increase in $L_s$, suggesting the existence of a characteristic length scale $\zeta$. The exponential fit function $C \exp(-L_s/\zeta)$ is shown as dashed lines. Here the fit parameter $C$ represents the limiting value of the relative yield strain in the limit of zero sub-system size, and $\zeta$ is the estimated value of the characteristic heterogeneity length scale intrinsic to each model system. {\color{black} This length scale $\zeta$ defines the minimum size of an amorphous solid region capable of undergoing a local failure within an elastic background, without being affected by other plastic events or boundary conditions}. In 2D BMLJ and 3D soft repulsive models, which have short-range interactions, our fit finds  $\zeta$ varies between $5a$ and $6a$. {\color{black} Interestingly, this is similar to values frequently considered in the literature concerning the local rheological measurements} \cite{2009TsamadosPRE, Puosi2015, barbot2018local} where it is interpreted as the length scale below which Hooke's law for linear, continuum elasticity breaks down. Here we have provided the rationality of this choice by directly extracting an intrinsic scale via the proposed soft matrix method. For 3D $SiO_2$ and 3D Metallic glass we find $\zeta$ to be $\approx 12 cm$ and $\approx 8 cm$ respectively. The higher values in these models are presumably due to the long-range nature of inter-particle interactions not present in Lennard-Jones-type models typically reported in the literature. Figure \ref{fig2} presents additional data obtained using the frozen method for the purpose of comparison. {Under the frozen matrix protocol, small sub-systems exhibit a significant stress increase due to the surrounding rigid matrix, an effect that becomes less pronounced as the sub-system size increases. Recently, efforts have been made to mitigate such boundary effects by introducing a buffer layer between the sub-system and the background \cite{rottler2023thawed}.} 

After extracting a characteristic length scale $\zeta$, we focus on studying the influence of the preparation protocol on $\zeta$. { Age dependence is investigated with the 2D BMLJ system ($L=90$, $N=10K$).}

\begin{figure}[htpb]
\centering
\includegraphics[width=0.95\linewidth]{Fig3.eps}
\caption{{\bf Age Dependence.} (a) Relative yield strain as a function of sub-system size for different ages of the initial samples (2D BMLJ). (inset) The same plot in log-linear scale for a selected few ages. Dashed lines represent fits to exponential functions using which characteristic lengths $\zeta$ are extracted. (b) Characteristic length $\zeta$ as a function of energy $U_{init}$ (energy of samples at zero strain) showing the change $\zeta$ with an increase in the samples' age. Note that $U_{init}$ decreases with an increase in age. The solid line is just a guide to the eye and the dashed line is drawn for $\zeta = 5a$.}
\label{fig3}
\end{figure}

\section{\label{Age}Age dependence on $\zeta$}

It is well known that the mechanical response varies dramatically with the increasing age of the amorphous solid samples \cite{varnik2004study, vasisht2020emergence, bhaumik2021role}. For example, a poorly-aged samples often shows a more homogeneous ductile-like yielding behavior, while a well-aged or a well-annealed samples exhibit heterogeneous brittle-like yielding via shear banding \cite{vasisht2020emergence, 2021RichardMRS}. Thus it is essential to understand how $\zeta$  depends on the samples' age or preparation history, which we present here for 2D BMLJ model. We study samples quenched by infinite cooling rate (obtained by energy minimization of configurations generated at different initial temperatures $T_W$ to instantaneous inherent structures) as well as the ones prepared via finite cooling rate $\Gamma$. Around $250$ samples were prepared at infinite to finite but higher $\Gamma$, $100$ samples at moderate $\Gamma$, and $50$ samples at the lowest $\Gamma$. We characterize the age of these samples from the inherent structure energy $U_{init}$, which is higher for poorly-aged samples and lower for well-aged samples. We then obtain the $\Delta \gamma_y (L_s)$ via the soft matrix protocol - see Fig. \ref{fig3} (a) main panel. Note that we have scaled the abscissa by the age-dependent fit parameter $C$ for a better comparison among different ages. We find that poorly-aged samples ($U_{init} \geq -3.68 \epsilon$) show little or no dependence on age, but with increasing age ($U_{init} \leq -3.75 \epsilon$), the relative yield strain exhibit a slower decay and hence $\zeta$ increases significantly. In the inset of Fig. \ref{fig3} (a), we show the main panel data in the log-linear scale to reveal the exponential behavior. The dependence of $\zeta$ on the age of the sample is shown in Fig. \ref{fig3} (b).  We see that the value $\zeta \approx 5.0a$ for poorly-aged samples, and systematically increases with an increase in age or decrease in inherent structure energy. Our analysis clearly indicates that any measure of local mechanical properties must account for the size of the local probing region, as it explicitly depends on the age of the material. Further investigation is underway to determine the functional form of the age dependence $\zeta$. 

{It is compelling to discuss the possible significance of this new length scale and its connection to other scales. The length scale obtained from the soft matrix method is a coarse-graining length to measure local mechanical properties. It is the scale above which a subregion yields like the bulk. The other length scales which might be relevant in this context are: (i) Point-To-Set (PTS) length scale \cite{yaida2016point} referred to as $\zeta_{PTS}$, (ii) core size (referred to as $\zeta_{CO}$) of Quasi-localized modes (QLMs) extracted using the normal mode analyses \cite{2020LernerPNAS}, and (iii) the typical distance between these QLM cores (defined as $\zeta_S$), which defines the degree of disorder \cite{2020LernerPNAS}. The PTS length scale is associated with increasing amorphous ordering with decreasing temperature or better annealing. $\zeta_{PTS}$ and $\zeta_S$ are expected to be the same length scale. On the other hand, the length scale associated with the core size of the plastic event computed using the QLM eigenvectors is different from these two length scales. The size of the QLM core is found to increase with a decrease in age. We believe that the density of QLM cores ($\mathcal{N}$) has a strong bearing on the mechanical properties of the solids; rather than the core size itself, as $\zeta_S \sim \mathcal{N}^{-1/d}$, where $d$ is the spatial dimension. The soft-matrix length scale can be understood as the length scale over which the correlation in the displacement field, the strain or the stress field due to a plastic event dies out, such that a continuum elastic background is sufficient to describe the far-field behaviour. With increasing annealing, the density of defect zones (QLM cores) decreases, as does the relative size. This leads to an increase in the length scale over which the displacement field emanating from a plastic event remains correlated. Thus, it is expected that the soft-matrix length scale will be related to $\zeta_{PTS}$ and $\zeta_S$ rather than $\zeta_{CO}$, and it will increase with increasing annealing rather than decrease. A systematic comparison between these length scales will be very important for a better understanding of how some of these structural length scales play a role in the plasticity of amorphous solids. See the discussion section for more discussion on the length scale.}
\begin{figure*}[htpb]
\centering
\includegraphics[width=.95\linewidth]{Fig4.eps}
\caption{{\bf Local mechanical properties: comparison of estimates via soft matrix and frozen matrix.} (a.) Average yield strain $\gamma_y$, obtained from soft Matrix (left panel) and frozen Matrix (right panel) methods, as a function of sub-system size $L_s$ for two different ages ($U_{init}=-3.6 \epsilon$ corresponding to poorly-aged states and $U_{init}=-3.78 \epsilon$ corresponding to the well-aged samples). Fit lines (dashed dark green line - exponential fit and solid lines - power law fit) show a cross-over from an exponential regime at small $L_s$ to a power law regime at large $L_s$. The cross-over length (indicated in (a.) by dotted vertical lines) varies from $\sim 12 a$ in the the poorly-aged samples to $\sim 18 a$ in the well-aged samples, as obtained via the soft matrix method, unlike in frozen matrix (see main text) (b.) (inset) Distribution of local yield stress threshold $P(X)$ where $X = \sigma_y - \sigma_0$ and (main panel) the corresponding cumulative distribution $F(X)$ for the two different ages discussed in (a), with left and right panels corresponding to measurements from soft Matrix and frozen Matrix respectively. The fit lines  are from the Weibull distribution (see main text). (c.) Comparison of the distribution of local storage modulus ($\mu$), obtained via the two methods, as marked, measured using the 3D soft-rep model (poorly-aged); the lines correspond to Gaussian fits. The bulk storage modulus is marked with an orange circle.}
\label{fig4}
\end{figure*}

\section{\label{secLLS}Local mechanical properties and comparison with estimates from frozen matrix method} 

{\color{black} 
We present a comparative analysis of our soft matrix method with the previously introduced frozen matrix method, using the 2D-BMLJ model ($L=320$, $N=125K$ ). While the frozen matrix method has provided valuable insights on how to estimate yield stress threshold distribution and also demonstrate how this threshold increases with the age of the amorphous solid, the estimated yield stress magnitudes are quite large \cite{patinet2016connecting, barbot2018local} and the bulk modulus \cite{Mizuno2013} is also overestimated. Further, the method fails in providing important detailed features such as providing information about optimal size of the probing region and the age-dependence on pseudo-gap exponent \cite{barbot2018local}. In this section, we demonstrate that the soft matrix method offers a more accurate characterization of local mechanical properties, addressing these limitations effectively. In Fig. \ref{fig4}, we present a comparison between the two methods by showing the dependence of yield strain on $L_s$, as well as local yield stress and local bulk modulus statistics, computed from samples of two distinctly different ages, viz. $U_{init} = -3.6 \epsilon$ and $U_{init} = -3.78 \epsilon$. In Fig. \ref{fig4} (a) we show the variation of average yield strain ($\gamma_y$) with $L_s$ on a log-log scale. Unlike in the frozen matrix, the age dependence is quite evident in the soft matrix. More interestingly, in the soft matrix, the dependence of $\gamma_y  (L_s)$ shows a distinct change from exponential to power law, with the power law exponent showing an age dependence as well. Previous numerical studies of amorphous systems with periodic boundary conditions have shown that the average yield strain $\left<\gamma_y\right> \sim L^{-b_0 d}$, where $d$ is the spatial dimension and $b_0=0.7$ \cite{karmakar2010statistical}. In 2D, this implies a scaling of $\left < \gamma_y \right > \sim L^{-1.4}$. For the poorly-aged samples, our data are consistent with this prediction (see left panel of Fig. \ref{fig4}(a)), where we estimate a power-law exponent of approximately $1.6$. In contrast, well-aged samples exhibit a much steeper dependence, with an estimated exponent close to $2.6$.  For comparison, the frozen matrix method yields an exponent around $2.4$ (see right panel of Fig. \ref{fig4}(a)), largely independent of the samples' age.  It is interesting to note that the value of the exponent for the well-aged samples, as estimated from the soft matrix method, is close to what is obtained for the frozen matrix method --  with aging, the rigidity of the amorphous state increases and thus the elastic boundary that we construct eventually tends towards a more rigid matrix which that and the obtained response therefore becomes similar to a frozen boundary. 
We note that the soft elastic confinement used in our method differs from the standard periodic boundary conditions employed in Ref. \cite{karmakar2010statistical}. This distinction may influence the observed power-law exponent, presenting an interesting direction for future study. Further, unlike in Ref.\cite{karmakar2010statistical} where measurements are done in the steady state, we probe the statistics of the first plastic event, which is influenced by the preparation history of the amorphous solid~\cite{lerner2018protocol}.  

Next we compute the local yield stress threshold (X) using both the methods and study the corresponding statistics. Here $X = \sigma^i_y - \sigma^i_0$, where $\sigma^i_y$ and $\sigma^i_0$ are the yield stress and zero strain stress values at the local region $i$, whose linear length scale is of the order of $\zeta$. In Fig.~\ref{fig4}(b), we show the distribution $P(X)$ (inset) as well as the cumulative distribution $F(X)$ (main panel). In the soft matrix, we compute $X$ using the $L_s$ values obtained from Fig.~\ref{fig3}(b), with $L_s = 6 a$ for the poorly-aged samples and $L_s = 8 a$ for the well-aged samples. For the frozen matrix, following previous literature \cite{ruan2022predicting}, we use $L_s = 10 a$ for both ages (see {\it SI} for frozen matrix F(X) for varying $L_s$). While both methods show a shift in the peak of $P(X)$ to higher values for the well-aged samples, the shift is more pronounced in the soft matrix and the strong variation with age is clearly evidenced in $F(X)$. At small values of $X$, $P(X)$ differs significantly between the two ages in the soft matrix case, a contrast not captured by the frozen matrix. 

We also compare, across the two methods, the local storage modulus  ($\mu$) computed from the slope of the local load curve. In 2D systems, these curves are strongly affected by fluctuations and are noisy at the system sizes used in this work. These issues are reduced in 3D systems and therefore we use the 3D soft repulsive system ($L=42$ and $N=97556$) for this analysis. In Fig. \ref{fig4} (c), we show the distribution of the local storage modulus, $P(\mu)$, computed for both the soft matrix and frozen matrix methods, along with Gaussian fits to the measured data. Unlike the frozen matrix method, which strongly perturbs the elasticity of the background, the soft matrix maintains an elastic background close to the bulk, resulting in the obtained $P(\mu$) centered around the bulk storage modulus (orange circle in Fig. \ref{fig4} (d)), unlike the frozen matrix method for which the distribution is far shifted to larger values.

\begin{figure*}[htpb]
\centering
\includegraphics[width=0.98\linewidth]{Fig5.eps}
\caption{{\bf Local yield stress statistics.} (a.) Cumulative local yield stress threshold distribution $F(X)$ for varying $L_s$ for poorly-aged samples ($U_{init}=-3.6 \epsilon$) and well-aged samples ($U_{init}=-3.78 \epsilon)$. (b.) F(X) and scaled P(X) (inset) on log-log axes show distinct two branches corresponding to two different ages as well as the power-law behaviour near the small yield stress threshold, with $\theta$ being the pseudogap exponent. The solid lines in the inset correspond to power law fits with exponent $0.5$ (black solid line) and  $0.85$ (red dashed line) (c.) The Weibull fit parameters, scale parameter $\alpha$ and shape parameter $\beta$, obtained from fitting F(X), as a function of $L_s$ for poorly-aged and well-aged samples. The pseudogap exponent $\theta$ is related to saturated value of  $\beta$ as $\beta$ = $1 + \theta$ (shown as blue dotted and dashed lines at 1.5 and 1.85 respectively). (d.) The Weibull fit parameters obtained from the frozen matrix are shown for comparison purposes.}
\label{fig5}
\end{figure*}

\section{\label{secMech}Local yield stress statistics and pseudogap exponent}

Finally, we present a detailed investigation of the statistics of the local yield stress measured via the soft matrix method. In Fig.~\ref{fig5}, we show the dependence of local yield stress statistics on the probing lengthscale $L_s$. In panel (a), the cumulative distribution function (CDF) $F(X)$ is shown for poorly-aged and well-aged samples. The age dependence is clearly evident in the range of threshold values obtained from our simulations, with lower values observed in poorly-aged samples compared to well-aged ones. The CDF data fits well to the Weibull distribution function $F(X) = 1 - \exp[-(X/\alpha)^{\beta}]$, where $\alpha$ is the scale parameter and $\beta$ is the shape parameter. An increase in $\alpha$ shifts the distribution toward higher yield thresholds, while a higher $\beta$ leads to a steeper rise in $F(X)$, indicative of increased brittleness \cite{jiang2011study, ruan2022predicting}. The extracted values of $\alpha$ and $\beta$ as functions of $L_s$ are shown in Fig. \ref{fig5}(c). The differences in $\alpha$ and $\beta$ between the poorly-aged and well-aged samples are apparent. The scale parameter $\alpha$ decreases with increasing $L_s$ before saturating, with distinct differences in the rate and magnitude of change between the two aging conditions. Notably, $\alpha$ remains consistently higher in the well-aged samples, as the yield stress thresholds are expected to increase with age. The $\alpha$ obtained using the frozen matrix method (see Fig. \ref{fig5}(d)) exhibits a similar trend with $L_s$, but its magnitude is approximately an order higher and shows no apparent dependence on age. Turning our attention to the shape parameter $\beta$, we observe very interesting features. For poorly-aged samples, $\beta$ remains nearly constant with $L_s$, saturating around $1.5$. The saturation length corresponds to the transition length from exponential to power-law behavior, which occurs around $12$. In contrast, the well-aged samples shows that within $10a < L_s <30a$, $\beta$ decreases towards a saturation value of $1.85$. The saturation length is around the transition length ($18$) from exponential to power-law behavior. These observations highlight the significance of the probing length scales revealed through the use of the soft matrix. The change in the saturation value of $\beta$ clearly indicate a systematic change in the yielding behavior with age. For comparison, we present the variation of $\beta$ with $L_s$ from the frozen matrix method (see Fig. \ref{fig5} (d)). For $L_s > 10a$, $\beta$ shows a monotonic decrease with an indication of age dependence consistent with previous study \cite{ruan2022predicting}. Interestingly, the saturation value is comparable to that of the soft matrix, although the corresponding saturation $L_s$ is approximately twice as large. We point out that the initial increase of $\beta$ with $L_s$ ($4a \leq L_s \leq 10a$) could be related insufficient sampling or the size is too small to capture the full variability of the material’s disorder. This feature is observed only in the well-aged samples within the soft matrix, but appears at both ages in the frozen matrix, highlighting the influence of rigid boundary effects on the local mechanical properties in the frozen matrix method. Further we note that saturation values of $\beta$, which are $1.5$ for the poorly-aged samples and $1.85$ for the well-aged samples, correspond to pseudogap exponents $\theta$ by the relation $\theta = \beta - 1$ \cite{karmakar2010statistical}. The pseudogap exponent $\theta$ is a key indicator of marginal stability in amorphous solids \cite{lin2014density, lin2016mean}. This exponent reflects how close a system is to mechanical failure, with lower $\theta$ value in poorly-aged samples indicating more weak spots and higher $\theta$ value in well-aged samples signaling greater stability. The pseudogap exponent $\theta$ can also be determined from the small-$X$ power-law behavior of $F(X)$, which is evident in the log-log plot of $F(X)$ and $P(X)$ shown in Fig.~\ref{fig5} (b). The $F(X)$ shown in the main panel clearly distinguishes between the two ages, with the data separating into two distinct branches. In the inset, we show $P(X)$ for a range of $L_s$ where $\beta$ remains constant, scaling $P(X)$ appropriately to eliminate the trivial influence of $\alpha$. While improved statistics are needed to directly extract $\theta$ from $P(X)$, a clear power-law behavior at small $X$ is evident. To demonstrate that the value of $\theta$ is at least consistent with, if not precisely matching, the corresponding $\beta$, we include power-law fits with exponents $0.5$ and $0.85$. The corresponding data from the frozen matrix is presented in the {\it SI}, showing that the distributions do not exhibit any age-dependent branching. Moreover, the small-$X$ behavior of $P(X)$ is too noisy to reliably extract $\theta$ directly. {In summary, our study demonstrates the existence of a coarse-graining length scale for measuring local mechanical properties in microscopic simulations, and reveals clear differences in the marginal stability of amorphous solids with different thermal histories.}

\section{\label{secSummary}Summary and Discussion}
{\color{black} In this work, we perform molecular simulations of four different models of amorphous solids to demonstrate the existence of a characteristic intrinsic length scale, $\zeta$, within which the material can undergo local yielding inside a purely elastic background, unaffected by other plastic events or boundary conditions. We find that $\zeta$ depends distinctly on the age of the material—being smaller in poorly-aged samples and increasing with aging. This lengthscale is revealed via a novel simulation protocol, a major refinement over existing methods \cite{Sollich2010, Mizuno2013, patinet2016connecting, barbot2018local}, which we term the soft matrix method. In this approach, a selected local region within a sheared amorphous solid undergoes unconstrained relaxation, while the surrounding material is restricted to affine deformation using additional harmonic restoring forces. Using this framework, we also analyze local mechanical properties and show that, compared to the frozen matrix method, local yield thresholds and elastic moduli are determined more accurately. {Finally, by examining the distribution of local yield thresholds in microscopic simulations, we demonstrate that the key statistical feature, the pseudogap exponent, exhibits a clear dependence on the material's age, both in the small-threshold power-law regime of the yield threshold distribution and in the large-length-scale variation of the Weibull shape parameter.} In conclusion, our findings demonstrate that the soft matrix method not only captures age dependent and length scale dependent mechanical behavior across a wide class of disordered systems, but also remains effective even at small system sizes, as the interface between the subsystem and background is treated in a more seamless and physically consistent manner.}

A natural question arising from our results concerns the origin of the characteristic length scale $\zeta$ in amorphous solids. 
The simplest interpretation is that $\zeta$ corresponds to the spatial scale over which local plastic relaxations are devoid of any interfacial effects; i.e. it captures the intrinsic scale over which the first localized irreversible event would occur in a bulk system under mechanical loading. Thereby,  $\zeta$ would correspond to the spatial extent beyond which interactions from other plastic events begin to interfere, thereby screening the expected power-law correlations, as suggested in a recent work \cite{2021LemaitrePRE}. However, without a direct comparison, it remains unclear whether the length scale $\zeta$ obtained via the soft matrix method is equivalent to other known length scales, though it is highly likely that it is closely related to the spatial extent of quasi-localized vibrational modes \cite{Baggiolli2021-PRL-defect, Wu2023-defect}. Earlier studies have examined the existence of an intrinsic static correlation length scale associated with amorphous order in supercooled liquids and glasses \cite{2015KarmakarRPP}. In the context of sheared amorphous solids, similar length scales have been identified through non-affine shear deformation protocols \cite{weeks2007short, 2012MosayebiJCP, 2014MosayebiPRL, vasisht2018rate, Nussinov2011-shearlength, Nussinov2013-shearlength}. Yet, a systematic comparison with static length scales derived from non-shear-based approaches—such as the point-to-set length scale \cite{2008BiroliNPhys}, finite-size scaling of relaxation times \cite{2009KarmakarPNAS}, and the scaling of minimum eigenvalues \cite{2012KarmakarPhysicaA}—is still lacking. These various measures are expected to be proportional in the supercooled liquid regime \cite{2013BiroliPRL}. Our work thus opens up a promising avenue to compare and understand these different length scales, offering insights into the fundamental nature of spatial correlations in amorphous materials.

Despite the aforementioned limitations, it is noteworthy that the characteristic length scale $\zeta$ obtained through the soft matrix method exhibits remarkable similarity to the coarse-graining length scales reported in various studies on elasto-plastic models. These models often define an appropriate coarse-graining scale, typically around $5$ in suitable units, to capture the propagation of plasticity in the medium using the Eshelby kernel \cite{2009TsamadosPRE, Puosi2015, richard2020predicting, patinet2016connecting, barbot2018local, 2021Castellanos-lengthscale, 2022Castellanos-lengthscale}. It is interesting to observe that in order to study the effect of ageing or annealing in these elasto-plastic models, systematically increasing the coarse-graining length scale is often required to achieve a comparable correspondence with microscopic simulations \cite{lerner2018protocol}. Hence, it appears that the length scale derived from the soft matrix method may serve as the desired length scale for the development of a robust mesoscopic elasto-plastic model capable of incorporating ageing or annealing effects in a more fundamental manner. Thus, this method provides a crucial step toward developing such models, and, more significantly, offers key insights into how aging governs the distribution of weak regions, drives the emergence of marginal stability, and shapes the microscopic origins of mechanical response in disordered solids.

Overall, our work provides valuable insights into the microscopic behavior of soft disordered solids, enhancing our understanding of their macroscopic properties and paving the way for improved mesoscale models required for material design.
}
\begin{acknowledgments}
\noindent SS, PC, SS and VVV acknowledge the generous financial support from NSM grant $\mathrm{DST/NSM/R\&D \_HPC\_Applications/2021/29}$ as well as computational time on PARAM Yukti computing facility. SK acknowledges the Swarna Jayanti Fellowship, grants DST/SJF/PSA01/2018-19, and SB/SFJ/2019-20/05. SS acknowledge the Faculty Initiation Grant PHY/FIG/100804 from IIT Roorkee and PARAM Ganga computational facility. VVV acknowledge the computational time on the Chandra and Madhava computing facility at IIT Palakkad. Further we would like to thank Srikanth Sastry and Peter Sollich for fruitful discussions.
\end{acknowledgments}

\appendix
\section*{Appendix A: Methods}
The four different models studied in this work are as follows. (1.) {\bf 2D BMLJ:} Binary Mixture (65:35) of Lennard-Jones particles is a common glass forming material \cite{2009BruningJPCM}. The interactions smoothly goes to zero at the large distance. Two different system size were considered, N=10K and 125K at a reduced density $\rho=1.2$. (2.) {\bf 3D Soft-rep:} A system of 10\% polydisperse soft spheres interacting via the WCA potential \cite{vasisht2020computational}, is a typical colloidal model in 3D. At a volume fraction of $\phi=0.7$, system size of N=1K, 5K, 10K and 100K considered. (3.) {\bf 3D SiO$_2$:} Most common molecular glass is modeled using the VSP modified BKS potential of 1:2 binary mixture of silicon and oxygen atoms \cite{2004VoivodPRE}. System sizes of $N=13824, 46656$ particles at a density of $\rho=2.8\, gm/cm^3$ are considered. (4.) {\bf 3D Metallic Glass:} A metallic glass system $Cu_{64.5}Zr_{35.5}$ alloy using the embedded atom model (EAM) \cite{mendelev2019development}. System sizes $N=50000, 100000$ particles are considered at zero pressure.

\subsection*{Sample preparation protocol} 
For 2D BMLJ (65:35) and 3D Soft-rep (polydisperse) systems, equilibrium NVT MD simulations were performed at $T=5.0 \,\epsilon/k_B$ with $\rho=1.2$ and $\phi=0.7$ respectively and equilibrated configurations were quenched to $T=0.01$ at cooling rates $\Gamma$ varying from $10^{-2}$ to $10^{-6}$. These configurations were then subjected to potential energy minimization protocol (FIRE or Conjugate gradient, CG). 
For 3D Silica, configurations were equilibrated by NVT MD simulations at $T \approx 3100\,K$ and $\rho= 2.8 gm/cm^3$ followed by infinite quench by potential energy minimization (CG). For 3D metallic glass, NPT MD simulations were performed at zero pressure and at temperatures $T=1100, 1200, 1300 \, K$ to equilibrate the system. Then configurations were quenched to $T=300K$ at cooling rates $\Gamma=10^{10}, 10^{11}, 10^{12}\, K/s$ followed by potential energy minimization by CG. In each model and cooling rate a minimum of 10 to a maximum of 250 samples were prepared. The shear stress $\sigma$ is computed from the off-diagonal term of the virial stress tensor \cite{vasisht2020computational}. 

\subsection*{Shearing (AQS) protocol} We follow Athermal Quasi-static Shear (AQS) deformation method \cite{maloney2006amorphous}. In this procedure, an elementary shear strain $\delta \gamma \sim \mathcal{O}(10^{-5} - 10^{-6})$ is applied to a sample by the uniform Affine deformation rule: $x_{i}^{'} = x_{i}^{'} + \delta \gamma \, y_{i}, \, y_{i}^{'} = y_{i}, \, z_{i}^{'} = z_{i}$ to all particles $i$. Following which the sample is relaxed using an energy minimization technique (FIRE or CG) before deforming again. Here we chose the $\delta \gamma$ such that in most of the samples prepared we found only a single plastic event. The maximum applied strain $\gamma_{max}$ was chosen such that at least one plastic event occurs within $\gamma_{max}$. In all our simulations, {\it x} is the flow or shearing direction, {\it y} is the gradient direction and {\it z} is the vorticity direction.

\subsection*{Plastic event or Yielding} To detect the plastic event or the yielding event, we measure the largest non-affine displacement in the system. The displacement is computed from two minimized configurations (separated by the elementary strain $\delta \gamma$ and defined as $\Delta z^2_{max} (\gamma) = \max [z_i (\gamma + \delta \gamma) - z_i (\gamma)]^2$ where $z_i(\gamma)$ is the coordinate of a particle $i$ in the direction perpendicular to shear (gradient direction in 2D and vorticity direction in 3D). Comparing the stress drop and the potential energy drop we put a threshold on $\Delta z^2_{max}$, which is model dependent. The threshold chosen is also verified against the back-shear protocol \cite{barbot2018local}. We find that this method provides an excellent balance between sensitivity of detection and computational efficiency. More details are provided in {\it SI}. 
\subsection*{Local yield statistics}
Using the soft matrix method, we analyze the distribution of local yield stress in a 2D BMLJ model system consisting of 125000 particles with a system size of $L=320$. We randomly sample regions of size $L_s$ (with area $L_s^2$) throughout the system. We vary the $L_s$ between $4a$ and $40a$. To have sufficient statistical sampling, for each $L_s$, we make sure that the cumulative sampled area is atleast equal to total area of the system.

\bibliography{biblio}

\clearpage
\newpage
\section*{Supplementary Information}
\subsection*{Detecting plastic events} In Fig. \ref{SIfig1} we show the evolution of the system's shear stress ($\sigma$), potential energy (U) as well as the Maximum displacement with applied strain $\gamma$, for 2D BMLJ. Here we have considered the displacement in Y since the shearing direction is along X. In 3D systems, we consider Z, the vorticity direction. Clearly the jumps in the $\sigma$ and U is signed in term of kinks in the displacement. In order to differentiate the signal from the background noise, we compute the distribution of these displacements and is shown in Fig. \ref{SIfig2}. It is quite evident that we have a bimodal distribution and we take conservative estimate for the threshold to recognize the plastic event as $0.01$. This threshold is robust for varying ages of the sample. 
Similar analysis are done for other models to obtain the appropriate thresholds.

\noindent{\bf Plastic events in subsystem with elastic background : } We perform the above analysis in the soft matrix as well wherein only a subsystem is allowed to have plastic events. Even though the bimodality of the displacement distribution is coarsened (see Fig. \ref{SIfig3}), the threshold for the plastic event is quite evident and recognize the events accurately (see Fig. \ref{SIfig4}). 

\subsection*{Choice of k value in the soft matrix} In the soft matrix method we allow the sub-system to relax completely without any constraint and the background is allowed relax affinely. This is achieved by addition of a spring potential with a sprint constant k. The choice of k is made such that (1.) no large displacement occur in the background but also (2.) do not alter the system's mechanical properties. In the Fig. \ref{SIfig5} we show (for 2D BMLJ) the maximum displacement measured in the subsystem (left) and the full system (right), including the subsystem. For k=2 and k=5, we find that relative large displacement are occurring in the surround and not in the subsystem. But with the increase in the k, these events in the surrounding are curbed.  In the Fig. \ref{SIfig6} we show curvature of the PE profile (vs. $\gamma$) before the plastic event, the storage module (from the slope of $\sigma$ vs $\gamma$ before the plastic event) as well as the pressure difference $\Delta P = P_{k} - P_{k=0}$ for varying k values. These quantities are computed for the full system with soft matrix background. We find that for a range of k values (k<=20), these measures do not change and are similar to bulk value (dashed line in Fig. \ref{SIfig6}). Hence $k=10$ is a good estimate which satisfy both the criterion defined above. Similar protocols are followed for other models. 

{\color{black} \subsection*{$L_s$ dependence on the local yield threshold statistics} In Fig.\ref{SIfig7} we show the scaled plot of the normalized cumulative distribution function from soft as well as frozen matrix methods for two differently aged samples \cite{ruan2022predicting}. The scaling helps in identifying the $L_s$ beyond which minimal finite-size effects are observed. In soft matrix, poorly-aged samples shows a clear data collapse for all the $L_s$ values. In the well-aged samples for $L_s \geq 10a$, minimal finite size effects are found. Unlike in soft matrix, the frozen matrix data show $L_s$ dependence in both ages due to the rigid boundary imposed in the method. Going to even larger system size might reduce such effects. In Fig. \ref{SIfig8} we show complete yield threshold statistics from frozen matrix method, for (a) poorly-aged and  (b) well-aged samples. The lack of age dependence is evident in the range of threshold values obtained from our simulations but is highlighted in Fig. \ref{SIfig8} (c), where we show that for $L_s < 10a$, F(X) do not differentiate between ages. In Fig. \ref{SIfig8} (d) we show the small-$X$ behavior of $F(X)$ and $P(X)$, which clearly illustrates the method's insensitivity to the samples' age.}

\begin{figure}[b]
\centering
\includegraphics[width=.9\linewidth]{SI-data/SIFig1.eps}
\caption{{\bf Evolution with strain.} Stress $\sigma$ and potential energy U evolve with applied strain. Jumps in these quantities observed as the system show local yielding due to plastic events, hence non-affine displacement captured by maximum displacement measure. This data is for 2D BMLJ model and poorly-aged samples}
\label{SIfig1}
\end{figure}
\clearpage
\begin{figure}[t]
\centering
\includegraphics[width=.9\linewidth]{SI-data/SIFig2.eps}
\caption{{\bf Maximum displacement distribution.} Top plot show the raw data of maximum displacement and bottom its distribution. This data is for 2D BMLJ model and poorly-aged samples}
\label{SIfig2}
\end{figure}

\begin{figure}[b]
\centering
\includegraphics[width=.9\linewidth]{SI-data/SIFig3.eps}
\caption{{\bf Maximum displacement distribution in the soft matrix.} Top plot show the raw data of maximum displacement and bottom its distribution, within the subsystem. This data is for 2D BMLJ model and poorly-aged samples with a subsystem size of $L_s=16a$.}
\label{SIfig3}
\end{figure}

\begin{figure}[h]
\centering
\includegraphics[width=.95\linewidth]{SI-data/SIFig4.eps}
\caption{{\bf  Evolution with strain in the sub-system.}  Stress $\sigma$ and potential energy U evolve with applied strain. Jumps in these quantities observed as the system show local yielding due to plastic events in the sub-system. This data is for 2D BMLJ model (poorly-aged) in the soft matrix of k=10 and $L_s=16a$.}
\label{SIfig4}
\end{figure}

\begin{figure}[h]
\centering
\includegraphics[width=.9\linewidth]{SI-data/SIFig6.eps}
\caption{{\bf Various measures of the full system with the soft matrix background (2D BMLJ).} For $k \leq20$, these measures do not vary substantially. The dotted line is the measure for unconstrained (without the soft matrix) full system. Here $\Delta P = P_{k} - P_{k=0}$ and is measure at small strain of $10\delta \gamma$}.
\label{SIfig6}
\end{figure}

\clearpage

\begin{figure*}[t]
\centering
    \includegraphics[width=.95\linewidth]{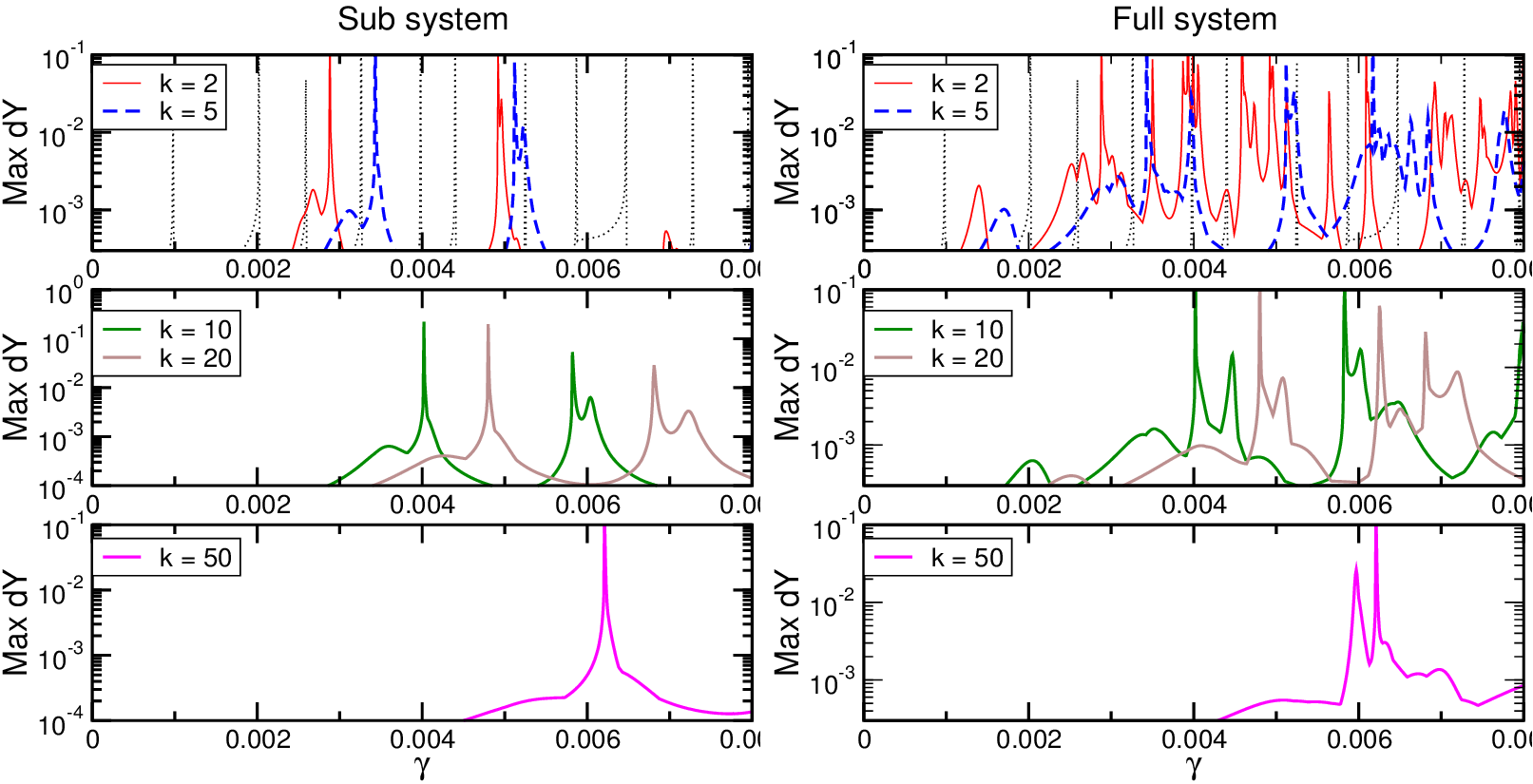}
\caption{{\bf k dependence on events inside and outside the sub-system.} Maximum displacement measure discussed in the text is computed for various k values, in the subsystem (left) and the full system (right) as a function of strain. The plastic event is signalled by the jump in the measure. Increasing the value of k, events occur only within the subsystem. This is for the 2D BMLJ system.}
\label{SIfig5}
\end{figure*}

\begin{figure*}[b]
\centering
\includegraphics[width=.8\linewidth]{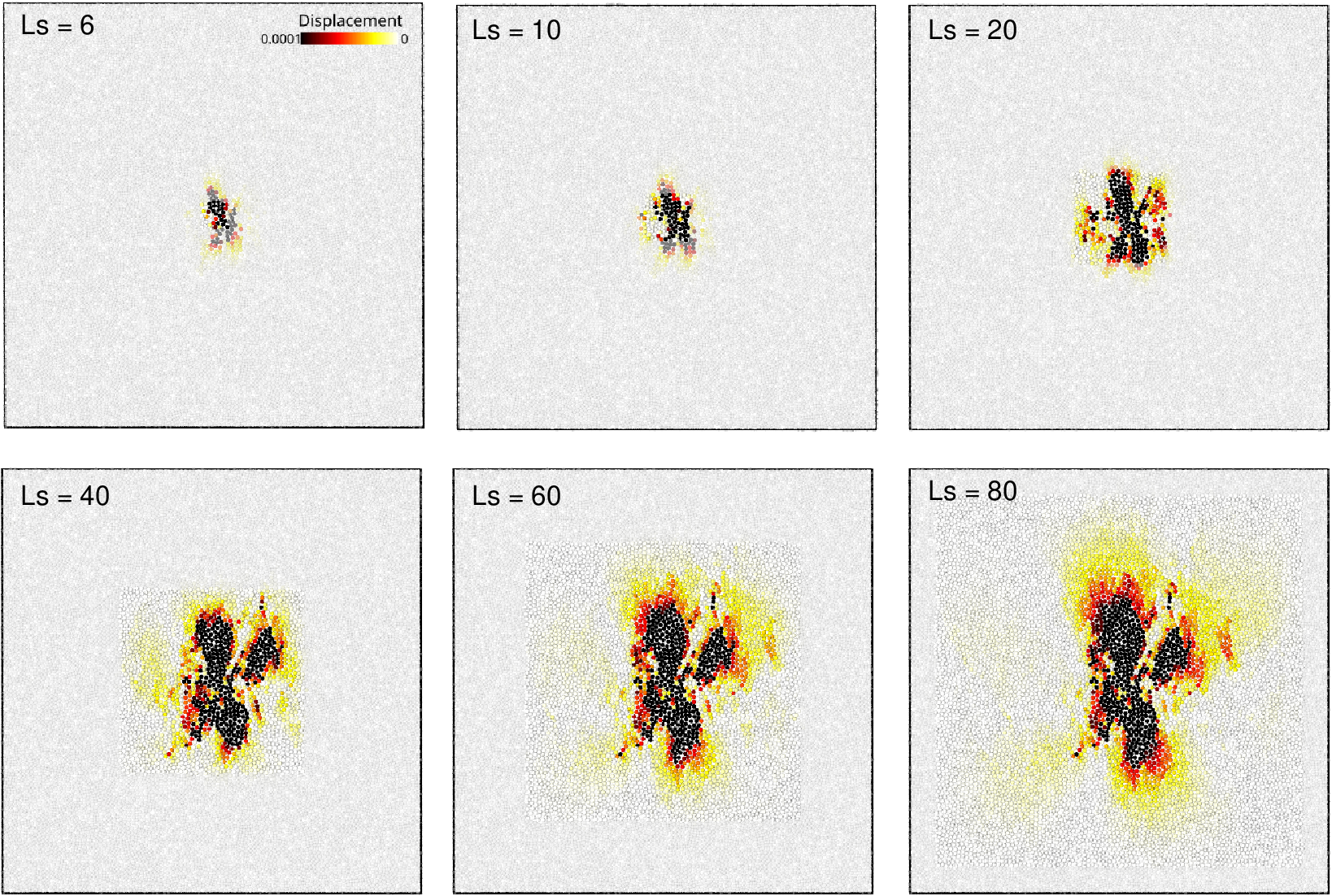}
\caption{{{\bf Displacement maps.} Obtained from 2D BMLJ simulation for fix $k = 10 \epsilon/a^2$ and varying Ls (6a, 10a, 20a, 40a, 60a, 80a). The heatmap coloring scheme is based on the particle displacements (white indicates no displacement and
black corresponds to the displacement of magnitude 0.0001). The region where soft matrix is imposed is made translucent
to emphasize the sub-system.}}
\label{SIfig9}
\end{figure*}

\begin{figure*}[t]
\centering
\includegraphics[width=.9\linewidth]{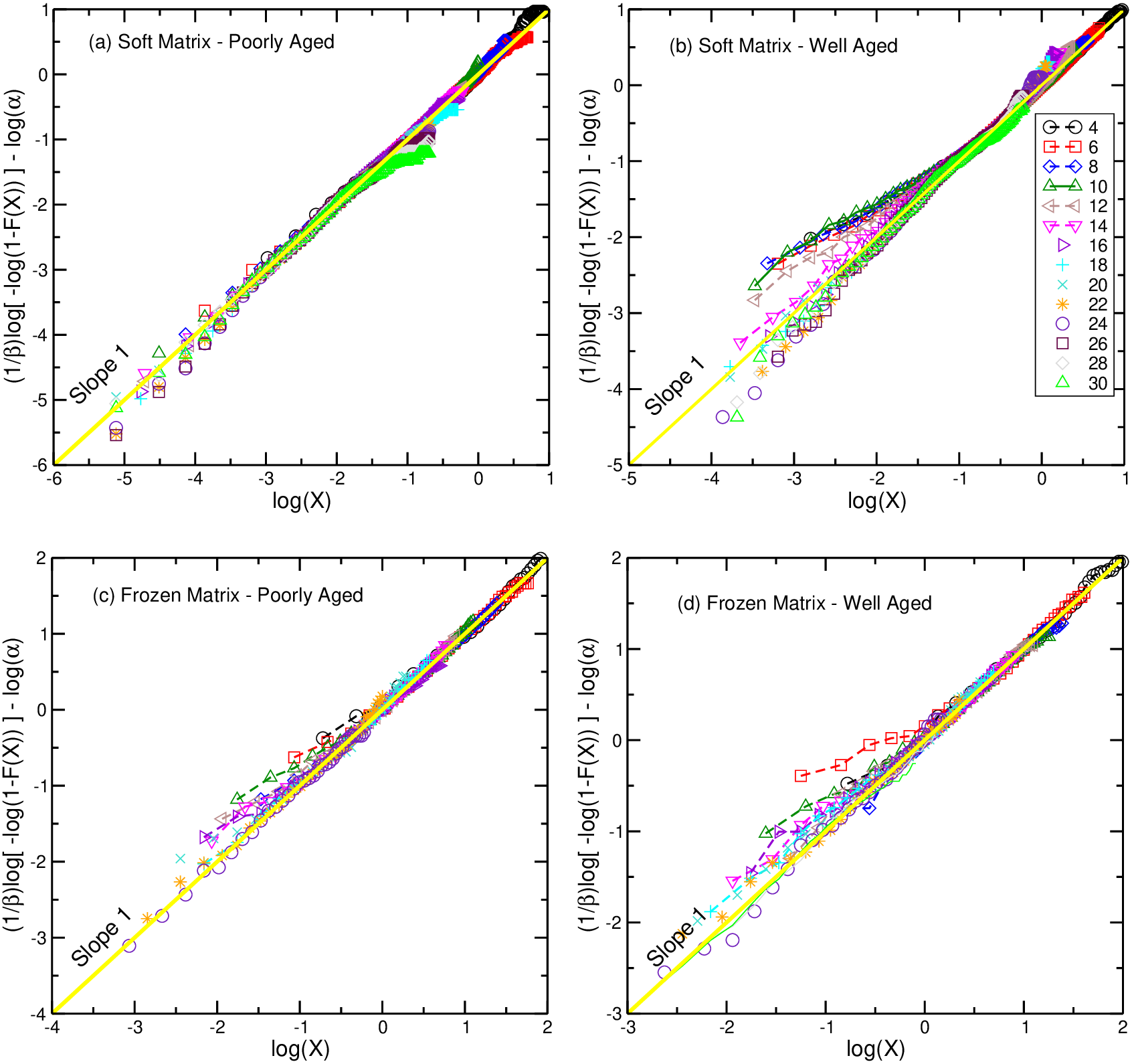}
\caption{{\bf Scaled local yield stress statistics.} The cumulative distribution function $F(X)$ is transformed to assess the validity of Weibull scaling and identify the minimum sampling size $L_s$ required for meaningful statistics. The transformation gives a linear relation and is highlighted by yellow solid line with slope $1$. In (a) and (b) we show the scaled $F(X)$ from soft matrix for poorly-aged and well-aged samples respectively. In (c) and (d) we show the scaled $F(X)$ from frozen matrix for poorly-aged and well-aged samples respectively. In soft matrix, for the complete range of $L_s$ varying from $4a$ and $40a$, poorly-aged samples shows a linear relation and in well-aged samples for $4a \leq L_s \leq 10a$ a finite probe size dependence is evident. In the same range, independent of age, the frozen matrix show a finite probe size depenence which is due to the rigid background.}.
\label{SIfig7}
\end{figure*}

\begin{figure*}[b]
\centering
\includegraphics[width=.95\linewidth]{SI-data/SIFig7.eps}
\caption{{\bf Local yield stress statistics from frozen matrix.} Cumulative local yield stress threshold distribution $F(X)$ for varying $L_s$ for (a) Poorly-aged samples ($U_{init}=-3.6 \epsilon$) (b.) Well-aged samples ($U_{init}=-3.78 \epsilon)$ . (c.) Comparison of F(X) across different ages, emphasizing the minimal variation observed for $L_s < 10a$ (d.) The Weibull fit parameters $\alpha$ and $\beta$. The scaling parameter ($\alpha$) shows little variation with age, and its absolute values, linked to the average yield threshold, are notably high, approximately an order of magnitude greater than those of the soft matrix. The shape parameter ($\beta$) exhibits some age dependence, although the saturation value seems to converge to a similar value. Large system simulations are required to confirm the same}.
\label{SIfig8}
\end{figure*}
\end{document}